**Published manuscript**

This manuscript is a pre-print version of the article available at OSA Publishing:
https://doi.org/10.1364/OL.44.005505

**Copyright**

**Citation text**

Christian R. Petersen, Mikkel B. Lotz, Getinet Woyessa, Amar N. Ghosh, Thibaut Sylvestre, Laurent Brilland, Johann Troles, Mogens H. Jakobsen, Rafael Taboryski, and Ole Bang, "Nanoimprinting and tapering of chalcogenide photonic crystal fibers for cascaded supercontinuum generation," Opt. Lett. **44**, 5505-5508 (2019)

# Nanoimprinting and Tapering of Chalcogenide Photonic Crystal Fibers for Cascaded Supercontinuum Generation

CHRISTIAN R. PETERSEN,[1,6,†,*] MIKKEL B. LOTZ,[2,†] GETINET WOYESSA,[1] AMAR N. GHOSH,[3] THIBAUT SYLVESTRE,[3] LAURENT BRILLAND,[4] JOHANN TROLES,[5] MOGENS H. JAKOBSEN,[2] RAFAEL TABORYSKI,[2] OLE BANG[1,6,7].

[1] DTU Fotonik, Technical University of Denmark, Kongens Lyngby DK–2800, Denmark.
[2] DTU Nanolab, Technical University of Denmark, Kongens Lyngby DK–2800, Denmark.
[3] Institut FEMTO–ST, CNRS, Université Bourgogne Franche–Comté UMR6174, Besançon, France
[4] SelenOptics, 263 Avenue du Gal Leclerc, Campus de Beaulieu, 35700 Rennes, France
[5] Glasses and Ceramics Group, ISCR UMR–CNRS 6226, University of Rennes 1, 35042 Rennes Cedex, France
[6] NORBLIS, Virumgade 35D, 2830 Virum, Denmark
[7] NKT Photonics A/S, Blokken 84, DK–3460 Birkerød, Denmark
*Corresponding author: chru@fotonik.dtu.dk , [†]These authors contributed equally

**Improved long–wavelength transmission and super–continuum (SC) generation is demonstrated by anti–reflective (AR) nanoimprinting and tapering of chalcogenide photonic crystal fibers (PCF). Using a SC source input spanning from 1–4.2 μm, the total transmission of a 15 μm core diameter PCF was improved from ~53 % to ~74 % by nanoimprinting of AR structures on both input– and output facets of the fiber. Through a combined effect of reduced reflection and red–shifting of the spectrum to 5 μm, the relative transmission of light >3.5 μm in the same fiber was increased by 60.2 %. Further extension of the spectrum to 8 μm was achieved using tapered fibers. The spectral broadening dynamics and output power was investigated using different taper parameters and pulse repetition rates.**

SC light sources based on chalcogenide glass optical fibers are among the main candidates for enabling new and improved mid–infrared photonic applications within broadband imaging and spectroscopy [1–3]. Currently, the most promising method for generating SC reaching far into the mid–infrared fingerprint region is through so–called cascaded generation using a series of concatenated silica, fluoride, and chalcogenide fibers that gradually extend the spectrum from the near–infrared to the mid–infrared [4–7].

One critical point in particular is the coupling to– and from the chalcogenide fiber. Due to the high refractive index of chalcogenide glasses, a significant part of the pump light is reflected at the end facets, which means that both the input– and output power is substantially reduced. One way to mitigate this is by using AR thin film coatings such as $Al_2O_3$ or diamond deposited on the fiber end facets [8]. However, deposition of thin films on fiber facets is an elaborate and time-consuming cleanroom process that requires careful sample preparation, and the resulting coating is typically narrowband, which is not ideal for cascaded generation with a broadband input source. Another way to achieve AR over a broad bandwidth is by direct thermal nanoimprinting [9–11]. Due to the low glass transition temperature ($T_g$) of chalcogenide glasses (~160–280 °C) the process can be performed using simple heating elements, and takes only a few minutes to complete.

Another critical aspect of coupling light to a chalcogenide fiber is avoiding damage when launching high peak power pulses. The simplest solution to allow for better power handling is to scale up the fiber core. However, when the core diameter is increased the nonlinearity of the fiber is reduced, thus hindering further extension of the spectrum. In order to increase the nonlinearity without sacrificing the damage threshold, the fiber diameter can be gradually reduced along the propagation direction through thermal tapering, thus improving the MIR generation efficiency [12,13].

In this work we investigate these two thermal post–processing techniques, nanoimprinting and tapering, in order to improve the output power and bandwidth of cascaded SC sources. We chose to work with single–material $Ge_{10}As_{22}Se_{68}$ PCF in order to avoid issues with varying $T_g$'s and viscosities between the core and cladding in step–index fibers (SIF). The pump source for the experiments was a SC source based on a master oscillator power amplifier (MOPA) configuration with a variable pulse repetition rate and 1 ns pulse envelope [14]. The MOPA was used to pump a 7 µm core diameter $ZrF_4$–$BaF_2$–$LaF_3$–$AlF_3$–$NaF$ (ZBLAN) glass fiber generating a spectrum covering 1–4.2 µm.

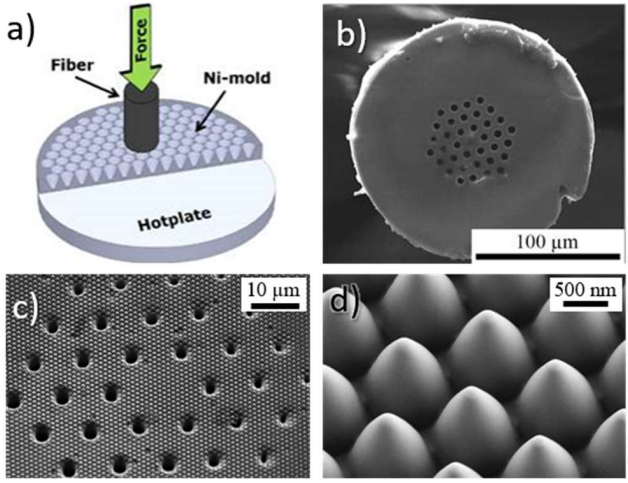

Fig. 1. (a) Principle of fiber nanoimprinting using a heated Ni mold. (b) SEM image of the PCF prior to imprinting. (c) SEM image of the imprinted core and air–clad region. (d) High magnification SEM of the imprinted structures in the core.

The principle of fiber nanoimprinting is depicted in Fig. 1(a). A nickel (Ni) mold with the negative structure is placed on a hot plate and heated to above the $T_g$ of the chalcogenide glass (∼175 °C [15]). For more details regarding the fabrication of the mold see refs. [9,16]. The fiber is then pressed against the mold for a brief period of time, after which the fiber is quickly released to allow for the imprinted structure to settle. Force was applied by manually moving a translation stage, and the combination of force, imprint time, and temperature enabled for varying degrees of structuring and deformation of the fiber. In general, higher temperatures allow for faster imprinting using low force, which also reduces the heat conduction to the rest of the fiber and thus reduces deformation.

Fig. 1(b) shows the PCF used in the experiments, which has a core diameter of 15 µm and cladding diameter of 170 µm. The best imprints were achieved using a temperature between 230–250 °C applying a very small translation force for just a few seconds. Fig. 1(c) and 1(d) show high magnification scanning electron microscope (SEM) images of the imprinted structures, which were designed to have a pitch of 1050 nm and a height of around 1350 nm, providing excellent anti-reflection from 2.6–6.2 µm (<4 % reflection) peaking at 4 µm (<1 % reflection) [9,16]. The size of the structure meant that for wavelengths below ∼2.2 µm the imprint acts as a diffraction grating, effectively scattering much of the light into the cladding. This effect is seen from the microscope image of Fig. 2(b), which shows the chalcogenide fiber lighting up near the butt–coupling interface with the ZBLAN fiber from the pump source. It is also apparent from the image that the imprinting procedure in this case caused significant deformation of the fiber tip, resulting in a so–called elephant's foot shape. However, this deformation affects mainly the outermost part of the cladding leaving the core area more or less intact except for some hole shrinkage, as is apparent from Fig. 1(c). The effect of fiber tip deformation and hole shrinking on the output beam profile is seen when comparing the beam profiles of Fig. 2(c) and 2(d), which shows the near fields of the non–imprinted and imprinted fibers, respectively. The near field images were captured using an uncooled PbSe camera covering from 1–5 µm wavelength (Tachyon 6400, NIT). In the non–imprinted fiber output beam the PCF structure is clearly seen, while the imprinted output beam shows a significant distortion of the air–cladding structure while maintaining a circular core–guided mode. Note that the intensity scale on the figures is logarithmic in order to clearly see the air–hole patterns. By optimizing the temperature, pressure, and imprinting time it was possible to reduce the fiber tip deformation while maintaining a full imprint of the structure, but it was found to have no influence on the transmission since the core area is seemingly unaffected by the deformation. To measure the output spectra the PCF output was butt–coupled to a 150 µm core diameter chalcogenide patch cable (IRF–Se–150, IR Flex) and guided to a grating–based spectrometer covering from 0.2–5 µm (Spectro 320, Instrument Systems). From comparing the red output spectrum of the non–imprinted fiber in Fig. 2(a) to the purple and blue output spectra of the 1–face (output) and 2–face imprinted fiber, respectively, it is clear that a significant fraction of the light below 2.2 µm is lost due to diffraction. Such properties could potentially be useful for reducing the linear and nonlinear absorption of shorter wavelengths, in particular two-photon absorption, which may improve the power handling. It is also evident from the spectra that the long–wavelength transmission was improved after imprinting. In order to quantify the improvement the total output power and output power after a 3.5 µm long–pass filter (LPF) was measured for the non–imprinted and the 2–face imprinted fibers. Many fiber samples with similar lengths around 20–30 cm were prepared and measured, and because the imprinting was performed manually the results varied from producing only minor or no improvement in transmission, to even reducing the output power compared to the non–imprinted reference. However, some samples also produced a significant increase in transmission, and these results are presented in Table 1. The total transmission improvement (ΔT) between the non–imprinted and 2–face imprinted fibers ranged from 7.0–27.9 % and the maximum total transmission (T) was 73.8 %. As seen from the orange curve of Fig. 2(a), a significant part of the pump spectrum is below the diffraction limit of the imprinted structure, so the transmission increase >3.5 μm was also measured and found to range from 12.3–60.2 % with a maximum transmission of 98.4 %. However, this extremely high value is attributed to a combination of both reduced reflection and increased red–shifting of wavelength components below 3.5 μm due to the higher transmission.

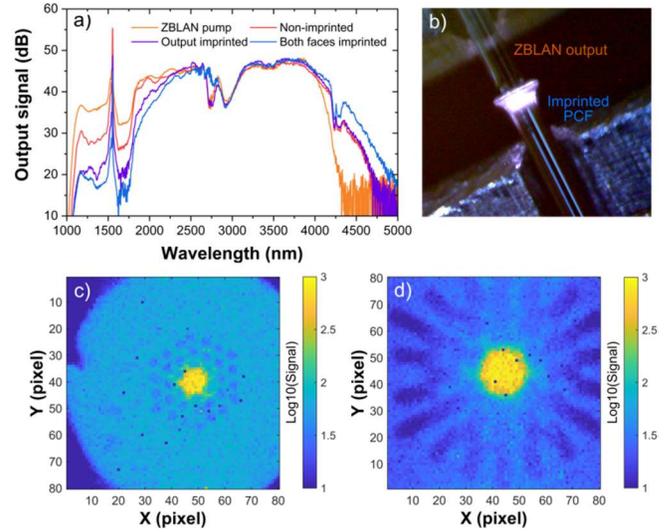

Fig. 2. (a) Comparison between the input spectrum (orange) and the non–imprinted (red), output imprinted (purple), and 2–face imprinted (blue) output spectra, respectively. (b) Microscope image of the ZBLAN to PCF butt–coupling revealing scattered light from diffraction caused by the imprinted structure. (c,d) Output beam profile of the non–imprinted (c) and imprinted (d) fiber. Note that log scale was used on the intensity axis to better visualize the air–cladding.

Table 1. Measured transmission (T=$P_{out}/P_{in}$) of non–imprinted and 2–face imprinted PCFs[a]

| Fiber sample | $P_{in}$ (mW) | $P_{out}$ (mW) | T (%) | ΔT[a] (%) |
|---|---|---|---|---|
| **Non #1** | 139.8 [31.9] | 74.4 [18.9] | 53.2 [59.2] | – – |
| **2–face #1** | 139.8 [31.9] | 83.5 [23.3] | 59.7 [73.0] | 10.9 [18.9] |
| **2–face #2** | 139.8 [31.9] | 86.5 [25.8] | 61.9 [80.9] | 14.1 [26.8] |
| **2–face #3** | 133.5 [31.2] | 98.5 [30.7] | 73.8 [98.4] | 27.9 [60.2] |
| **Non #2** | 160.5 [– –] | 72.5 [19.9] | 45.2 [– –] | – – |
| **2–face #4** | 160.5 [– –] | 78.0 [22.7] | 48.6 [– –] | 7.0 [12.3][a] |

[a] The improvement in transmission is defined as ΔT = T(2–face)/T(non). Square brackets denote values >3.5 μm.
[b] Calculated using the output power >3.5 μm for Non #2 and 2–face #4.

Previous work with fiber nanoimprinting has focused on large-core multi-mode $As_2S_3$ SIFs (50-100 μm core diameters) [11], and so it was not clear whether the micron scale structures and especially the fiber tip deformation would affect transmission of smaller core fibers. In that respect, this work represents the first demonstration of AR nanoimprinting of a single-mode chalcogenide PCF, achieving increased transmission in a fiber suitable for SC generation without significantly affecting the core mode of the fiber. Despite the increased transmission and red–shifting in the imprinted fibers the spectral extent was limited to 5 μm, so in order to increase the bandwidth even further, fiber tapering was for the first time to the authors' knowledge investigated for cascaded SCG. In a previous study on SC in tapered chalcogenide PCFs [12] using direct pumping it was found that the length of uniform fiber before ($L_b$) and after ($L_a$) the tapered fiber section ($L_t$) greatly influenced the bandwidth obtained, but the influence of the tapered core diameter was never fully explored. Here we have investigated the influence of the tapered core diameter ($d_c$) in five different tapered fibers, whose parameters are summarized in Table 2.

Table 2. Taper parameters.

| Taper ID | $d_c$ (μm) | $L_b$ (cm) | $L_t$ (cm) | $L_a$ (cm) | $L_{tot}$ (cm) |
|---|---|---|---|---|---|
| **A** | 6 | 6 | 11 | 18 | 35 |
| **B** | 6.5 | 10 | 15 | 11 | 36 |
| **C** | 7 | 2 | 12 | 19 | 33 |
| **D** | 7.5 | 9 | 15 | 18 | 42 |
| **E** | 6 | 27 | 30 | 12 | 69 |

All tapers were designed to have around 2.5 cm adiabatic taper transitions. When the fiber core is reduced the zero–dispersion wavelength (ZDW) is shifted towards shorter wavelengths from around 4.9 μm at the initial 15 μm diameter to around 3.6–4 μm at 6–7.5 μm diameter, respectively [12]. But while the reduced ZDW allow for more pump power to couple into the anomalous dispersion regime for improved SCG efficiency, the confinement loss edge is also pushed towards shorter wavelengths, which introduces a trade–off between efficiency and long–wavelength loss. Fig. 3(a) shows the resulting spectra from pumping tapers A–E with around 270 mW of average power at 1 MHz pulse repetition rate. The spectra were measured using a fiber–coupled Fourier–tranform infrared (FTIR) spectrometer (FOSS, custom) covering from 2.8–14 μm. The total transmission of the non-imprinted tapers A–D varied from 31–38 %, with the highest output power of 103 mW achieved for taper D.

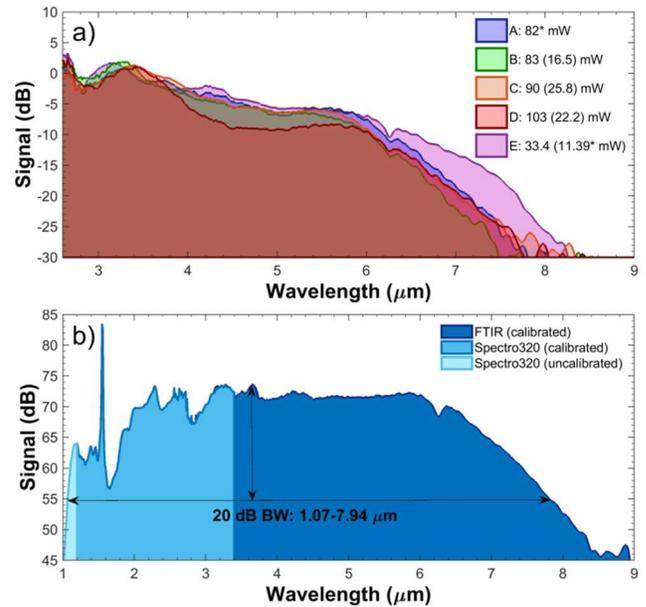

Fig. 3. (a) FTIR spectra of the long–wavelength edge obtained for tapers A–E for a pump power of ~270 mW. The corresponding measured total output power and power >3.5 μm (brackets) is indicated in the legend (*=estimated). (b) Stitched total output spectrum for taper E using 350 mW pump power and generating a 41 mW continuum with a 20 dB bandwidth from 1.07–7.94 μm.

Despite the differences in taper parameters, tapers A–C produced very similar output spectra reaching up to around 8 μm at the –30 dB level, while taper D was limited to 7.5 μm and exhibited significantly lower flatness compared to the three others. For a 7.5 µm core diameter the ZDW is estimated to be around 4 µm, which is exactly where we see a dip in the output spectrum of taper D. Such a dip followed by a peak or pedestal of lower intensity is commonly observed when pumping in the normal dispersion region close to the ZDW [17,18]. Since the SC pump starts to cut off at around 4 µm the solitons coupled into the PCF will initially experience normal dispersion and therefore only the part that broadens into the anomalous dispersion via self–phase modulation (SPM) will then contribute to the long–wavelength edge. As a result the amount of power >3.5 µm is only ~21 % in taper D (22.2 mW) compared to ~29 % in taper C (25.8 mW).

To investigate whether the spectral extension was limited by the taper length, another taper denoted Taper E was tested. Taper E has the same 6 µm core diameter as taper A, but twice the tapered section length and a longer length of uniform fiber before the taper. It is seen in Fig. 3(a) that the longer taper improves the long–wavelength generation, but reduces the output power significantly due to increased losses. Fig. 3(b) shows the entire taper E output spectrum at ~350 mW pump power (41 mW output power), where maximum broadening was observed covering from 1.07–7.94 µm at the −20 dB level. Further increasing the pump power did not improve the long–wavelength edge, as seen from the trend in Fig. 4(a), which shows that for these pump and taper parameters the broadening is limited to around 8 µm.

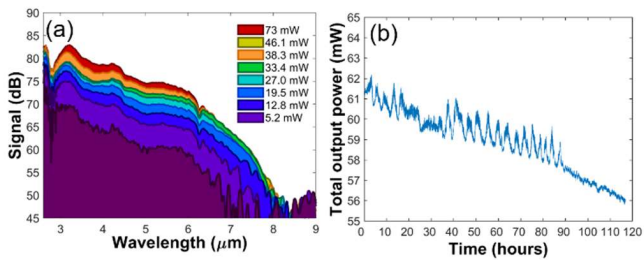

Fig. 4. (a) Spectral broadening in taper E with increasing output power (brackets: power >3.5 µm). (b) Output power degradation in Taper C during 116.5 hour testing.

This behavior was found in all of the tapers tested, which suggests that further broadening is limited mainly by the soliton dynamics of the MOPA source and ZBLAN fiber rather than confinement loss or dispersion in the tapers [5,7]. Lastly, the output power stability was tested. Note that at the point of this measurement (4 months later) taper C had degraded significantly due to exposure to ambient lab conditions [15], and so the obtainable bandwidth and output power was reduced. The insert shows the degradation in output power during a 116.5 hour continuous power stability measurement, and adjusting the coupling did not improve the output power. This highlights the need for proper sealing of the optical facets, which could for instance be realized by either fully collapsing the air–holes, or by splicing an unstructured fiber end–cap onto the PCF prior to imprinting.

In conclusion, we have demonstrated for the first time improved transmission and broadening in cascaded SC generation through nanoimprinting and tapering of $Ge_{10}As_{22}Se_{68}$ chalcogenide PCFs. The maximum total transmission of a nanoimprinted uniform fiber was increased by 27.9 % relative to the non–imprinted fiber, and the average power >3.5 µm was increased by up to 60.2 %. By tapering a non-imprinted 15 µm core PCF down to a core diameter of 7 µm over a 12 cm section a spectrum covering up to ~8 µm with 90 mW output power and 25.8 mW >3.5 µm was achieved. By combining the two post–processing methods we expect that the output power can be increased to more than 120 mW with more than 40 mW above 3.5 µm without damaging the fiber. Such power levels would enable applications such as optical coherence tomography to push further into the mid–IR for improved inspection of ceramics, coatings, polymers, and fiber–reinforced composites [19].

**Funding.** European Union H2020 (722380, 732968); Innovation Fund Denmark (4107–00011A).